\documentclass[prl,aps,amsmath,twocolumn,floatfix,showpacs]{revtex4}
\usepackage{epsfig}
\usepackage{bm}

\begin{document}

\title{Robust state preparation of a single trapped ion by adiabatic passage}

\author{Chr. Wunderlich, Th. Hannemann}
\affiliation{National University of Ireland, Maynooth, Co. Kildare,
Ireland \footnote{Present address: University of Siegen, Fachbereich
Physik, 57068-Siegen, Germany}}

\author{T. K\"orber, H. H\"affner, Ch. Roos, W. H\"ansel, R. Blatt}
\affiliation{Institut f{\"u}r Experimentalphysik, Universit{\"a}t
Innsbruck, Technikerstrasse 25, A-6020 Innsbruck, Austria}

\author{F. Schmidt-Kaler}
\affiliation{Institut f{\"u}r Quanteninformationsverarbeitung,
Universit{\"a}t Ulm, Einstein 11, D-89069 Ulm, Germany}

\date{\today}

\begin{abstract}
We report adiabatic passage experiments with a single trapped
$^{40}$Ca$^+$ ion. By applying a frequency chirped laser pulse with
a Gaussian amplitude envelope we reach a transfer efficiency of
0.990(10) on an optical transition from the electronic ground state
S$_{1/2}$ to the metastable state D$_{5/2}$. This transfer method is
shown to be insensitive to the accurate setting of laser parameters,
and therefore is suitable as a robust tool for ion based quantum
computing.
\end{abstract}


%03.67.Lx Quantum computation
%32.80.Qk Coherent control of atomic interactions with photons

\maketitle

It is the interplay between different technologies that is
stimulating novel developments aiming at the ambitious goal of a
future large-scale quantum computer \cite{ChuangBuch}. As recent
research has shown, considerable promise lies in the application of
nuclear magnetic resonance (NMR) technology to ion-trap based
quantum computing \cite{CHILDS2000,GULDE03}. While ion based quantum
computing has strong assets concerning the preparation of
multi-particle entangled states \cite{SACKETT00,ROOS2004} and the
highly efficient readout of qubit states using projective
measurements \cite{RIEBE2004,BARRETT2004,Wunderlich03}, liquid state
NMR quantum computing relies on well developed radio frequency (rf)
techniques which have enabled the most complex
\cite{CUMMINS2003,JONES2003,VANDER2004} sequences of quantum logic
gate operations to date with about 10$^{2}$ to 10$^{3}$ rf-pulses
\cite{VANDER2001}.

The basic construction principle of an elementary quantum computer
with trapped ions relies on linear cold ion crystals serving as
quantum register. Two of each of the ions' electronic states serve
to store elementary bits of quantum information (qubits) which are
coherently manipulated by the application of laser \cite{CIR95} or
microwave pulses \cite{Mintert01} with well defined timing,
frequency and phase. With a number of operations applied, on single
ions individually or on groups of ions a quantum algorithm may be
implemented.

Composite gate operations \cite{CHILDS2000}, initially developed in
the context of NMR experiments, have already enabled complex tasks
in ion traps like the demonstration of quantum teleportation
\cite{RIEBE2004,BARRETT2004}, which comprises about 30 laser pulses
of different frequency, phase and amplitude. In order to further
increase the complexity of algorithms and to improve the robustness
of single and multiqubit quantum logic gates, all parameters
characterizing the electromagnetic field driving qubit transitions
have to be freely adjustable, thus allowing for the implementation
of pulses with arbitrary amplitude and phase envelope. For this
purpose, a suitable waveform having these characteristics is
digitally generated in the rf-domain and then mapped
phase-coherently onto a fixed frequency laser or microwave field for
qubit manipulation. Here, as a first application we demonstrate
robust adiabatic passage (RAP) in a single trapped ion qubit system.

%Here we use a digitally synthesized rf-waveform with which we can
%freely program all pulse parameters used for qubit manipulation.
%Thus arbitrarily shaped and chirped pulses can be implemented.

In this publication first we briefly review some elements of the
theory of rapid adiabatic passage (RAP), then give a short
description of the experimental setup which allows the generation of
complex laser pulses. Subsequently, experimental data demonstarting
RAP are compared to the expected outcome. Finally, we sketch a
number of possible applications of the RAP method as elements of a
toolbox for ion based quantum computing.

\begin{figure}
\epsfxsize=1\linewidth \epsfbox{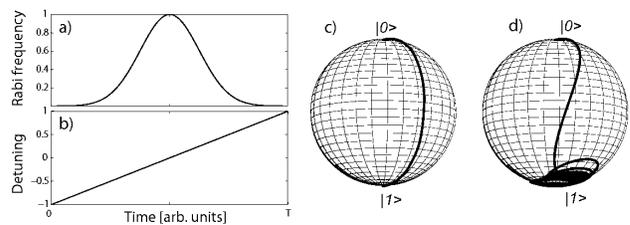}
% \setlength{\epsfxsize}{\columnwidth} \epsffile{fig1.eps}\par
% \vspace{2ex}\par \caption[rf pulse]
\caption[rf pulse] {Sketch of the temporal variation of the
rf-waveform to form the laser pulse used for the experiments. The
amplitude envelope of the light field is proportional to the Rabi
frequency and hs a Gaussian shape. The envelope is plotted in units
of the maximal Rabi frequency obtained during the laser pulse (a).
The linear frequency chirp (b) is plotted units of $\Delta\nu$, the
frequency range over which the detuning $\delta$ is swept during an
individual laser pulse. For the experimental studies, we vary the
pulses by variation of the peak laser intensity, thus changing the
Rabi frequency, and by variation of $\Delta\nu$, the chosen
frequency range of the detuning. c) Pictorial illustration of the
RAP method using the Bloch sphere representation of the atomic
two-level system. With $\Delta \nu$ = 400~kHz the evolution Bloch
vector evolution is fully adiabatic and results in perfect transfer.
d) Same as c), however with $\Delta\nu=1400$ kHz. Here, the
population transfer fails to be adiabatic. The parameters used for
the simulations shown in c) and d) are identical to those in the
experimental situation of fig.~2.} \label{fig:theory}
\end{figure}

For the theoretical description of RAP we model the atom as a
two-level system with quantum states $|0\rangle$ and $|1\rangle$ and
use a frequency-chirped laser light field coupling both levels. The
interaction sweeps through the resonance $\nu_{atom} =
\omega_{atom}/ (2\pi) $ of the two levels. It is convenient to
define a detuning rate $\Delta \nu / T$, where $\Delta \nu$ is the
frequency difference of start and stop frequency of the field
driving the atomic transition and $T$ the time duration of the
chirp. During the chirp, the light field amplitude is smoothly
turned on and off with a Gaussian envelope, proportional to
$\exp(-(t-T/2)^2/(2\sigma^2))$ with $\sigma=T/(6\sqrt{2})$  and
truncated at times $t=0$ and $t=T$, see Fig.~\ref{fig:theory}~(a)
and (b). The maximum of the light field amplitude is reached at the
frequency $\nu_{atom}$. The resulting atomic 2-level dynamics is
easily modeled using optical Bloch equations. A qualitative and
intuitive picture can be derived if we consider a Bloch sphere
representation of our system. Here the state vector
\begin{align}\label{statevector}
    \psi=c_0 |0\rangle + c_1 |1\rangle
\end{align}
is written as a vector $\mathbf{R}$ with components
$R_x=c_0c_1^*+c_0^*c_1,R_y=i(c_0c_1^*-c_0^*c_1),R_z=|c_0|^2-|c_1|^2$.
Likewise, an interaction $\Omega$ is written as vector
$\mathbf{\Omega}$=$(Re(H_{01})$, Im$(H_{01}), \delta)$, where
$H_{01}$ is the interaction Hamiltonian (in a frame rotating with
angular frequency $\omega_{atom}$) where $\hbar\omega_{atom}$
denotes the energy separation between levels $|0\rangle$ and
$|1\rangle$ and $\delta$ the detuning of the driving field from the
atomic resonance. The long lifetime of both atomic levels and the
fast dynamics allows us to neglect spontaneous decay and dephasing
\cite{ITANOV2005}. Under these conditions the equation of motion for
the Bloch vector dynamics can then be formulated \cite{Feynman57}
simply as
$\dot{\mathbf{R}(t)}=\mathbf{\Omega}(t)\times\mathbf{R}(t)$, in
obvious analogy with magnetic resonance phenomena. First
demonstrations of RAP in ensembles have been in magnetic resonance
phenomena \cite{Treacy68}, and in a variety of optical applications,
see \cite{Broers92,Wanner98,BERGMANN98} and also references therein.

If we start off with a weak electro-magnetic field tuned below
resonance and with the atom in $|0\rangle$, $\mathbf{\Omega}$ will
be almost aligned with the Bloch vector at the south pole of the
sphere. Because both vectors point nearly in the same direction, the
resulting Bloch nutations are of small amplitude. By increasing the
frequency of the driving field towards resonance, thus reducing
$\delta$, and simultaneously increasing the interaction strength,
$\mathbf{\Omega}$ moves toward the equator of the Bloch sphere. If
this is done slowly enough the Bloch vector $\mathbf{R}$
adiabatically follows $\mathbf{\Omega}$. At the equator we start to
lower the interaction strength again and continue to increase the
driving frequency, until $\mathbf{\Omega}$ and the Bloch vector
point toward the north pole: adiabatic transfer has taken place. The
corresponding trajectory of the Bloch vector during RAP is
numerically evaluated and shown in Fig. 1c). As long the
adiabaticity condition is fulfilled, with a temporal change
$|\mathbf{\dot{\Omega}}|/|\mathbf{\Omega}| \ll |\mathbf{\Omega}|$,
the transfer is robust and can be achieved over a broad range of
parameters. The limits of adiabaticity are illustrated by a second
Bloch nutation, with the passage through resonance scanned too
rapidly (i.e., $\Delta\nu/T$ is too large), see
Fig.~\ref{fig:theory}~(d). Residual nutations about the state
$|1\rangle$ are clearly visible and the final population of state $
|1\rangle$ at the end of the laser pulse depends strongly on the
exact duration and intensity of the pulse.

For the experiments, a single $^{40}$Ca$^+$ ion is stored in the
effective harmonic potential of a linear Paul trap. For a detailed
description of the experimental setup we refer the reader to
\cite{SCHM03}. Under typical operating conditions we observe axial
and radial motional frequencies $(\omega_\text{ax},
\omega_\text{rad})= 2\pi$~(1.2, 5.0)\,MHz, respectively. The ion is
Doppler-cooled on the S$_{1/2}$ to P$_{1/2}$ transition near 397~nm
to a mean phonon number of ($\bar{n}_{radial}, \bar{n}_{axial})$ =
(5, 15). The electronic level S$_{1/2}, m_j=-1/2$ is identified with
$|0\rangle$ and D$_{5/2}, m_j=-1/2$ with $|1\rangle$, respectively.
For the coherent manipulation on this transition, we modulate the
output of a Ti:Sapphire laser with an acousto-optical modulator
(AOM) in double-pass configuration. The radio frequencies and phases
that are applied to the AOM transfer directly to the light field
\footnote{Due to the double-pass configuration, the modulation of
laser frequency and phase is twice the applied rf-modulation. The
gaussian chirped laser field waveform is revealed from the
rf-waveform by the independently measured AOM transfer function.}.

The complex rf-waveform is provided by a novel versatile frequency
generator (VFG) that is based on a field-programmable gate array.
Important specifications are: a frequency range from 0 to 150~MHz,
an amplitude resolution of 16 bit, the option to lock to an external
clock (in our case a Rb-atomic clock), a switching time of 5~ns,
phase continuous or phase coherent switching between arbitrarily
many frequencies, multiple frequencies output, and an interface that
is programmable via a USB bus using a personal computer. Without the
typical restrictions due to a limited memory, in this device the
amplitude, phase and frequency of arbitrarily many shaped rf-pulses
are freely programmable.

To match the required AOM frequency near 230~MHz, we mixed the
output of the VFG near 80~MHz with the 150~MHz signal from an
rf-synthesizer \footnote{Marconi Inc., Signal gen. 2019A}. Unwanted
frequency components (e.g. near 70~MHz) from the mixer are rejected
by the AOM \footnote{Brimrose Inc.,TEF-270-100} due to its limited
bandwidth.

In the experiment, for a demonstration of RAP transfer, we apply a
temporal sequence to a single ion: a) The ion is prepared initially
in $|0\rangle $, b) a RAP transfer pulse $|0\rangle \rightarrow
|1\rangle$ is applied and c)  the excited state population
$P_{|1\rangle}$ is detected by an electron shelving technique
\cite{DEHMELT75}. This is accomplished by driving the $S_{1/2} -
P_{1/2}$ and $D_{3/2} - P_{1/2}$ dipole transitions with laser light
near 397 and 866~nm and monitoring the blue fluorescence emitted by
the ion with a photomultiplier. The internal state of the ion is
discriminated with an efficiency close to 100$\%$.

\begin{figure}
\epsfxsize=1\linewidth \epsfbox{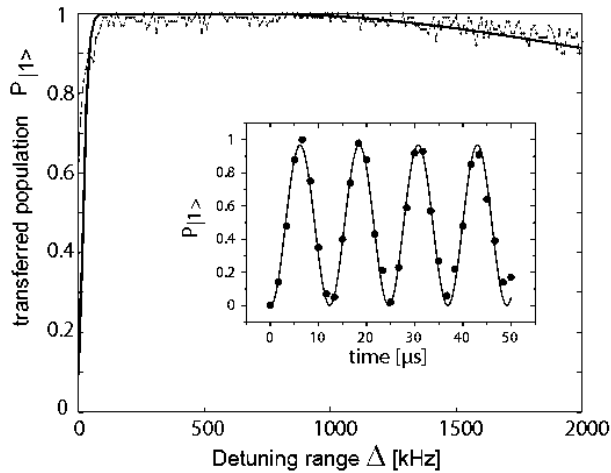} \caption[results] {Adiabatic
transfer $|0\rangle \rightarrow |1\rangle$. The excited state
population $P_{|1\rangle}$ is plotted for different rates of change
of the laser detuning $\delta$ during the pulse, that is for
different values of $\Delta\nu / T$ with $T$ fixed at 150~$\mu$s.
The transfer works efficiently in the wide range between $\Delta\nu
=$ 200~kHz and $\Delta\nu =$ 500~kHz, with average efficiency of
99.0(1.0)~\%. The solid line represents the result of an atomic
two-level calculation for $\Omega_{Rabi}$ = 512~kHz. Inset: The Rabi
frequency is directly determined by observing resonant Rabi
oscillations. From a sine fit to this data we obtain $\Omega_{Rabi}$
with a fractional error of 0.5~\%. Thus, all parameters for the
simulation are independently fixed, and the solid line is not fitted
to the data but results entirely from the model.} \label{fig:exp}
\end{figure}

Relevant parameters for RAP such as the Rabi frequency and the
detuning rate are varied in step (b). Keeping the time $T$ =
150~$\mu$s constant, the range of detuning $\Delta\nu$ is varied.
The results are plotted in Fig.~\ref{fig:exp}: The main experimental
result is the wide plateau of efficient population transfer from
state $|0\rangle$ to state $|1\rangle$. For $\Delta\nu$ ranging from
200~kHz to 500~kHz 99.0~(1.0)\% of the population is transferred
from the initial state to the target state. Even for the larger
range of detunings from 100~kHz to 600~kHz, almost perfect transfer
of 98.7(1.1)~$\%$ is achieved. The two-state simulation predicts a
transfer efficiency close to 100\% for the plateau in
Fig.~\ref{fig:exp}. Imperfect state preparation and read-out are
expected to limit RAP.

It is this  robustness, that should make RAP well suited for pulsed
Raman ground state cooling: For ground state cooling, a pulse of
laser radiation near the red vibrational sideband at
$\nu_{laser}=\nu_{E_{|1\rangle}} - \nu_{E_{|0\rangle}} - \nu_{trap}$
is applied to a single ion or a linear crystal of ions e.g. in an
experimental situation like \cite{KING98} for $^{9}Be^+$ or for
ground state cooling of a crystal of $^{43}Ca^+$ ions using two
Raman beams near 395~nm. Here, the cooling rate suffers from the
fact that the sideband Rabi frequency depends on the vibrational
quantum number as $\Omega_{n \rightarrow n-1} \propto \sqrt{n-1}$.
Thus, with the relevant cooling transition not driven equally strong
in case of a thermal distribution of vibrational states, the cooling
rate is decreased and even trapping states might occur at certain
vibrational quantum states $n$ if the interaction corresponds to an
integer $2\pi$-cycle \cite{MORIGI99,BLATT95}. Experimentalists have
adapted the duration of the Raman pulses to optimize the scheme. RAP
cooling, however, completely transfers {\em all} vibrational states
from $|0,n\rangle \rightarrow |1,n-1\rangle$ such that the
subsequent radiative decay closes the cooling cycle. Involuntarily,
in conventional pulsed Raman cooling the carrier transition is
weakly excited which leads to heating processes. In contrast, the
RAP cooling avoids this problem, as the carrier is excited but
de-excited again during the adiabatic ramp.

RAP might be used also for the efficient read-out of qubits in
systems with ground state qubits, such as hyperfine states of
$^{43}$Ca$^+$, F=4 to F=3, or the Zeeman levels $S_{1/2}, m_j = \pm
1$ in $^{40}$Ca$^+$. In both cases, a Raman transition can be used
for coherent qubit manipulation, but detection of the qubit logic
state does not work, as both qubit states would scatter photons
under excitation on the dipole transitions near 397~nm and 866~nm.
For electron shelving detection one of the qubit levels has to be
transferred to the $D_{5/2}$ state which then appears dark under
excitation on the dipole transitions. We propose for this transfer a
RAP pulse to achieve a robust and reliable read-out. The
alternative, a resonant $\pi$-pulse, would require a technically
demanding laser frequency stabilization and pulse area calibration.
For quantum gate operations techniques similar to RAP have been
proposed recently \cite{SAN2005}.

Further applications are expected in cavity QED schemes. Neutral
atoms have been used to demonstrate a deterministic single photon
emission \cite{HENNRICH2003} and similar schemes are currently
investigated for the case of trapped ions coupled to an optical
cavity. The emission of deterministic photons relying of RAP may
improve the feasibility.

In conclusion, we have shown RAP on a single trapped ion.
Experimental data and numerical simulation are in good agreement.
The experiments show that the $\pi$-transfer pulse is robust and
well suited for quantum logic operations and improved optical
cooling techniques.

{\em Acknowledgment} Ch. W. and Th. H. acknowledge financial support
from Science Foundation Ireland under grant number 03/IN3/I397 and
the Deutsche Forschungsgemeinschaft. T. K. acknowledges financial
support by the FWF. We acknowledge support by the European
commission (QGates).

\end{document}